\documentclass[10pt, conference]{IEEEtran}

\IEEEoverridecommandlockouts

\usepackage{amsmath}
\usepackage{graphicx}
\usepackage{subfigure}
\usepackage{curves}
\usepackage{amsfonts}
\usepackage{amsmath,epsfig}
\usepackage{stfloats}
\usepackage{amssymb}
\usepackage{cite}
\usepackage{algorithm}
\usepackage{algorithmic}
\usepackage{epstopdf}

\begin{document}

\title{Polar Coding for Block Fading Channels}
\author{\IEEEauthorblockN{Mengfan~Zheng, Meixia~Tao, Wen~Chen}
	\IEEEauthorblockA{Department of Electronic Engineering,\\Shanghai Jiao Tong University, Shanghai, P. R. China\\Emails: \{zhengmengfan, mxtao, wenchen\}@sjtu.edu.cn}
	\and
    \IEEEauthorblockN{Cong~Ling}
    \IEEEauthorblockA{Department of Electrical and Electronic Engineering,\\Imperial College London, United Kingdom\\Email: c.ling@imperial.ac.uk}
	

}

\maketitle

\begin{abstract}
 In this paper, we consider the problem of polar coding for block fading channels, with emphasis on those with instantaneous channel state information (CSI) at neither the transmitter nor the receiver. Our approach is to decompose a block fading channel of $T_c$ symbols per coherent interval into $T_c$ binary-input sub-channels in a capacity-preserving way, and design a polar code for each of them. For the case when instantaneous CSI is available at the receiver, a random interleaver can be used to enable joint encoding and decoding of all sub-channels so as to enhance the finite length performance. It is shown that our proposed schemes achieve the ergodic capacity of binary-input block fading channels under various CSI assumptions.
\end{abstract}


\section{Introduction}

The fading channel is a widely adopted time-varying model for real-world wireless communications. In this model, the channel gain changes over time satisfying a certain distribution, called the channel distribution information (CDI). In a block fading channel model, the channel gain is assumed to be constant over a fixed time interval $T_c$, known as the coherent time, and change to a new independent value afterwards. In many of today's communication systems, channel estimation is performed in the first place to obtain the instantaneous channel state information (CSI), and then data transmission follows. However, in many scenarios, the coherence time is very short (e.g., only a few symbol intervals). In this case, channel estimation may significantly reduces the overall data rate. Besides, the estimation precision is quite limited. Consequently, communication without instantaneous CSI (or noncoherent communication) is preferable.

Polar codes are the first family of codes that provably achieves the capacity of any binary-input symmetric memoryless channels with low encoding and decoding complexity \cite{arikan2009channel}. Later, polar codes are generalized to asymmetric channels while still capacity-achieving \cite{honda2013asymmetric}. There have been studies on polar coding for fading channels under various CSI assumptions. In \cite{boutros2013fading}, polar coding for quasi-static fading channels with two states was studied. Polar coding for block fading channels with full CSI and i.i.d. fading channels with CDI was considered in \cite{Santos2013rayleigh}. For block fading binary symmetric and
additive exponential noise channels with CSI at the receiver (CSI-R), a hierarchical polar coding scheme was proposed in \cite{si2014fading}, which achieves capacity, but only works for block fading channels with finite states. A simple method for construction of polar codes for Rayleigh fading channel was presented in \cite{trifonov2015rayleigh}. Polar codes and polar
lattices for i.i.d. fading channels with CSI-R were constructed in \cite{liu2016fading}, which achieve the ergodic capacity through single-stage polarization. All of the aforementioned polar coding schemes for block fading channels require the coherent time to be very large. As far as we know, polar coding for block fading channels with arbitrary finite coherent time has not been investigated in literature yet.

By viewing transmitted symbols in a coherent block as a supersymbol, we can design polar codes for block fading channels by using techniques in coded modulation, i.e., multilevel coding (MLC) \cite{Imai1977mlc,Wachsmann1999mlc} and bit-interleaved coded modulation (BICM) \cite{Caire1998bicm,Guilleni2008bicm}. Polar coded modulation, with both the MLC approach and the BICM approach, has been investigated, e.g., \cite{Seidl2013pcm,Mahdavifar2016bicm}. It is shown that the MLC-based polar coded modulation scheme achieves the code modulation capacity, while the BICM-based scheme suffers a certain rate loss.

In this paper, we aim to design capacity-achieving polar codes for block fading channels, especially for those with only CDI. A block fading channel with coherent time $T_c$ can be decomposed into $T_c$ parallel sub-channels. In the case when only CDI is known by the communicators, the sub-channels are correlated, and we use an MLC-based approach to design polar codes since such an approach preserves channel capacity. When instantaneous CSI is available at the receiver side or at both sides, the sub-channels become independent from each other. Thus, we can use an interleaver to randomize symbols from different sub-channels, and perform joint encoding and decoding so as to enhance the finite length performance. We refer to this scheme as the BICM-based approach in this paper. We show that in all the above mentioned cases, our proposed schemes achieve the ergodic capacity of any binary-input memoryless block fading channel.

The rest of this paper is organized as follows. In Section \ref{S:problemstate} we introduce the block fading channel model and describe the main idea of our schemes. Section \ref{S:PolarPri} provides some related knowledge on polar codes. Details of our proposed schemes are presented in Section \ref{S:scheme}. Section \ref{S:Numerical} shows some numerical results on achievable rates of block Rayleigh fading channels and the synthesized sub-channels in our scheme. Section \ref{S:discuss} concludes this paper with some discussions.

\section{Problem Statement}
\label{S:problemstate}

We consider a block fading channel model with coherent interval $T_c$. At time interval $i$ ($i=1,2,...$), the channel is modeled as
\begin{equation}
\mathbf{y}^i=h^i\mathbf{x}^i+\mathbf{w}^i,
\end{equation}
where $h^i\in \mathbb{R}$ is the channel gain at time interval $i$, $\mathbf{x}^i=[x^i_1,x^i_2,...,x^i_{T_c}]^{\mathrm{T}}\in\{-1,1\}^{T_c}$ is the binary input signal after BPSK modulation, $\mathbf{y}^i=[y^i_1,y^i_2,...,y^i_{T_c}]^{\mathrm{T}}\in \mathbb{R}^{T_c}$ is the channel output, and $\mathbf{w}^i=[w^i_1,w^i_2,...,w^i_{T_c}]^{\mathrm{T}}\in \mathbb{R}^{T_c}$ is the white Gaussian noise, with $w^i_j\sim \mathcal{N}(0,\sigma^2)$ for every $j\in[T_c]$. Let $X_j$ and $Y_j$ ($j\in[T_c]$) respectively be the random variables standing for the $j$th input and output symbols of a coherent block, and $H$ be the random variable for the channel gain.

First, let us discuss the case when both the transmitter and the receiver only have the CDI of the channel. Consider a series of transmissions over $N$ fading blocks. In this paper, we call the $N$ consecutive coded blocks a \textit{frame}. Denote $\mathbf{X}=[\mathbf{x}^1,...,\mathbf{x}^N]$ and  $\mathbf{Y}=[\mathbf{y}^1,...,\mathbf{y}^N]$, and let $\mathbf{z}_j$ ($j\in[T_c]$) denote the $j$th row vector of $\mathbf{X}$. Then the mutual information of a transmission frame can be expanded as
\begin{align}
I(\mathbf{X};\mathbf{Y})
&=\sum_{j=1}^{T_c}I(\mathbf{z}_j;\mathbf{Y}|\mathbf{z}_{1:j-1}),\label{expansion}
\end{align}
where $\mathbf{z}_{1:j-1}$ is short for $\{\mathbf{z}_1,...,\mathbf{z}_{j-1}\}$. Similar abbreviations will be used throughout this paper. Note that
\begin{align}
\lim\limits_{N\rightarrow\infty}\frac{1}{N}I(\mathbf{z}_j;\mathbf{Y}|\mathbf{z}_{1:j-1})&=I(X_j;Y_{1:T_c}|X_{1:j-1})\nonumber\\
&=I(X_j;Y_{1:T_c},X_{1:j-1}),\nonumber
\end{align}
which is the mutual information of a binary-input channel
\begin{equation}
W^{(j)}(\mathbf{y},x_{1:j-1}|x_j):\{-1,1\} \rightarrow \mathbb{R}^{T_c}\times\{-1,1\}^{j-1}, \label{subchannel}
\end{equation}
with the $j$th input symbol of a block being the input, and the whole output $\mathbf{y}$ together with the previous $j-1$ input symbols of the block being the output. 

Based on this expansion, we can use an MLC-based approach to design polar codes for block fading channels with only CDI. The encoding of a frame consists of $T_c$ component polar codes, designed for each of the $T_c$ sub-channels respectively. When an encoded frame is generated, the sender transmits it block by block. Having received a signal frame, the receiver uses a multistage decoder to decode the component polar codes one by one. At stage $j$ ($j\in[T_c]$), it decodes the $j$th sub-channel based on the received frame together with the estimates of previous stages. If the component polar codes are capacity-achieving, the ergodic CDI capacity of the binary-input block fading channel is also achievable with this scheme. 

For the case when only the receiver knows or both the transmitter and the receiver know the instantaneous CSI, the conditional mutual information of a transmission frame can be expanded as
\begin{align}
I(\mathbf{X};\mathbf{Y}|\mathbf{h})
&=\sum_{j=1}^{T_c}I(\mathbf{z}_j;\mathbf{y}_j|\mathbf{h}),\label{CSIR-2}
\end{align}
where $\mathbf{h}=[h^1,...,h^N]^{\mathrm{T}}$ is the channel state vector for the $N$ blocks, and $\mathbf{y}_j$ is the $j$th row of $\mathbf{Y}$. Equation (\ref{CSIR-2}) can be proved by the fact that when $\mathbf{h}$ is known, $\mathbf{y}_j$ is only related to $\mathbf{z}_j$ and independent from $\mathbf{z}_i$ ($i\in[T_c]\setminus j$). Details of the proof are omitted in this paper due to space limitation.

Consider $\mathbf{z}_j$ and $\mathbf{y}_j$ respectively as the input and output of a synthesized channel $W^{(j)}$. Obviously $W^{(j)}$ is an i.i.d. fading channel with the same CDI as the block fading channel, the ergodic capacity of which under the CSI-R assumption is $I(X_j,Y_j|H)$, i.e.,  $\lim\limits_{N\rightarrow\infty}\frac{1}{N}I(\mathbf{z}_j;\mathbf{y}_j|\mathbf{h})=I(X_j,Y_j|H)$. Since $W^{(1)},...,W^{(T_c)}$ are equivalent to each other in this case, we can see that the block fading channel is equivalent to $T_c$ i.i.d. fading channels with identical instantaneous CSI. Thus, simply designing a CSI-R polar coding scheme \cite{liu2016fading}, which treats the channel gain as another channel output, for each of the sub-channels independently is sufficient to achieve capacity. Nevertheless, the transmitter can actually use an interleaver to merge the parallel channels into a single one so as to improve the finite length performance. We will refer to this approach as the BICM-based scheme. Suppose the transmitter randomly permutes the positions of symbols in a frame with an interleaver. By deinterleaving the received frame, the signals can be seen as being transmitted through an i.i.d. fading channel with the same CDI as the block fading channel \cite{tse2005fundamentals}, provided that the interleaver is perfect and $N$ is sufficiently large. Since the CDI of the merged i.i.d. fading channel is the same as the original one, the ergodic CSI-R capacity is also the same. Thus, a polar code designed for the merged channel can achieve the ergodic capacity of the block fading channel.

\section{Preliminaries on Polar Codes}
\label{S:PolarPri}

First, we fix some notations that will be used in the sequel. $N=2^n$ with $n$ being an arbitrary integer. $\mathbf{G}_N=\mathbf{B}_N \textbf{F}^{\otimes n}$ is the generator matrix of polar codes, with $\mathbf{B}_N$ being the bit-reversal matrix and $\textbf{F}=
\begin{bmatrix}
1 & 0 \\
1 & 1
\end{bmatrix}$. The Bhattacharyya parameter $Z(X|Y)$ of a random variable pair $(X,Y)$ is defined as 
\begin{equation}
  Z(X|Y)=2\sum_{y\in \mathcal{Y}} P_Y(y)\sqrt{P_{X|Y}(0|y)P_{X|Y}(1|y)},
\end{equation}
with $X$ being binary and $Y$ being defined on an arbitrary discrete alphabet.

Let $X^{1:N}$ be $N$ independent copies of a binary random variable $X$, and $U^{1:N}=X^{1:N}\mathbf{G}_N$. It is shown that as $N$ goes to infinity, $U^{1:N}$ polarizes in the sense that $U^i$ ($i\in [N]$) becomes either almost independent of $U^{1:i-1}$ and uniformly distributed, or almost determined by $U^{1:i-1}$ \cite{arikan2010source}. Based on this phenomenon, for $\delta_N=2^{-N^\beta}$ with $\beta \in (0,1/2)$, we define the \textit{high entropy set} as
\begin{align}
\mathcal{H}^{(N)}_X&=\{i\in [N]:Z(U^i|U^{1:i-1})\geq 1-\delta_N\},\label{HX}
\end{align}
which satisfies
\begin{equation}
\begin{aligned}
\lim_{N\rightarrow \infty}\frac{1}{N}|\mathcal{H}^{(N)}_X|&=H(X).
\end{aligned}
\end{equation}

Let $(X,Y)\sim p_{X,Y}$ be a random variable pair with $X$ being binary and $Y$ being defined on an arbitrary countable set. Consider $X$ as the source to be compressed and $Y$ as side information of $X$. Let $U^{1:N}=X^{1:N}\mathbf{G}_N$. Similar to the single source case, conditioned on $Y^{1:N}$, $U^{1:N}$ polarizes as $N$ goes to infinity. For $\delta_N=2^{-N^\beta}$ with $\beta \in (0,1/2)$, define the \textit{reliable set} as
\begin{align}
\mathcal{L}^{(N)}_{X|Y}&=\{i\in [N]:Z(U^i|Y^{1:N},U^{1:i-1})\leq \delta_N\},\label{LXY}
\end{align}
which satisfy
\begin{equation}
\begin{aligned}
\lim_{N\rightarrow \infty}\frac{1}{N}|\mathcal{L}^{(N)}_{X|Y}|&=1-H(X|Y).
\end{aligned}
\end{equation}


Consider $N$ independent uses of a binary-input discrete memoryless channel $W(Y|X)$. Let $U^{1:N}=X^{1:N}\mathbf{G}_N$, and define $\mathcal{H}^{(N)}_X$ as in (\ref{HX}) and $\mathcal{L}^{(N)}_{X|Y}$ as in (\ref{LXY}). To construct a polar code for $W$, we define \cite{honda2013asymmetric}
\begin{align}
\mathcal{I}&\triangleq \mathcal{H}_X^{(N)}\cap \mathcal{L}_{X|Y}^{(N)}, \label{PC-I}\\
\mathcal{F}^r&\triangleq \mathcal{H}_X^{(N)}\cap (\mathcal{L}_{X|Y}^{(N)})^C, \label{PC-Fr}\\
\mathcal{F}^d&\triangleq (\mathcal{H}_X^{(N)})^C,\label{PC-Fd}
\end{align}
where $\mathcal{A}^C$ denotes the complement set of $\mathcal{A}$. The encoding is done by assigning $\{u^i\}_{i\in \mathcal{I}}$ with information bits and $\{u^i\}_{i\in \mathcal{F}^r}$ with uniformly distributed frozen bits (shared between the sender and the receiver), calculating $\{u^i\}_{i\in \mathcal{F}^d}$ with
\begin{equation}
\label{DetBit}
u^i= \arg\max_{u=\{0,1\}} P_{U^i|U^{1:i-1}}(u|u^{1:i-1}),
\end{equation}
and finally computing $x^{1:N}=u^{1:N}\mathbf{G}_N$ since $\mathbf{G}_N=\mathbf{G}_N^{-1}$.

Upon receiving $y^{1:N}$, the receiver computes an estimate $\bar{u}^{1:N}$ of $u^{1:N}$ as
\begin{equation}
\bar{u}^{i}=
\begin{cases}
u^i,\text{~~~~~~~~~~~~~~~~~~~~~~~~~~~~~~~~~~~~~~~~~~~~if } i\in \mathcal{F}^{r}\\
\arg\max_{u\in\{0,1\}}P_{U^{i}|U^{1:i-1}}(u|u^{1:i-1}),\text{~~~~if } i\in \mathcal{F}^{d}\\
\arg\max_{u\in\{0,1\}}P_{U^{i}|Y^{1:N}U^{1:i-1}}(u|y^{1:N},u^{1:i-1}),\\
\text{~~~~~~~~~~~~~~~~~~~~~~~~~~~~~~~~~~~~~~~~~~~~~~~~~if } i\in \mathcal{I}
\end{cases}.
\end{equation}

The rate of such a scheme, $R=|\mathcal{I}|/N$, satisfies
\begin{equation}
\lim_{N\rightarrow \infty}R=I(X;Y).
\end{equation}
The block error probability of such a scheme can be upper bounded by
\begin{equation}
\label{SC-EP}
P_e \leq\sum_{i\in\mathcal{I}}{Z(U^i|Y^{1:N},U^{1:i-1})}=O(2^{-N^\beta}).
\end{equation}

\section{Proposed Polar Coding Schemes}
\label{S:scheme}
\subsection{Only CDI Available}
The joint transition probability density function (PDF) of a coherent block without instantaneous CSI is given by
\begin{equation}
\label{transp}
p(\mathbf{y}|\mathbf{x})=\int_{0}^{\infty} \big{(}\prod_{j=1}^{T_c}p(y_j|h,x_j)\big{)}f(h) \mathrm{d}h,
\end{equation}
where $\mathbf{x}=(x_1,...,x_{T_c})$, $\mathbf{y}=(y_1,...,y_{T_c})$, $p(y_j|h,x_j)$ is the PDF for a given channel gain $h$, and $f(h)$ is the distribution of $h$. The transition PDF of the $j$th ($j\in[T_c]$) sub-channel is
\begin{equation}
p(\mathbf{y},x_{1:j-1}|x_j)=\sum_{x_{j+1:T_c}}p(x_{1:j-1},x_{j+1:T_c})p(\mathbf{y}|\mathbf{x}).\label{subchpdf}
\end{equation}

One may easily verify that the sub-channels are symmetric if the original fading channel is symmetric. In the symmetric case, uniform input distribution achieves the capacity, and the deterministic set defined in (\ref{PC-Fd}) is null. If the original channel is asymmetric, then the capacity-achieving input distribution may not be uniform distribution, and the deterministic set will not be empty.

For the $j$th ($j\in[T_c]$) sub-channel $W^{(j)}(\mathbf{y},x_{1:j-1}|x_j)$, let $\mathbf{u}_j=\mathbf{z}_j\mathbf{G}_N$. Define the high entropy set $\mathcal{H}^{(N)}_{X_j}$ in the same way as (\ref{HX}), and the reliable set $\mathcal{L}^{(N)}_{X_j|Y_{1:T_c}X_{1:j-1}}$ by
\begin{equation}
\begin{aligned}
\mathcal{L}^{(N)}_{X_j|Y_{1:T_c}X_{1:j-1}}&\triangleq\{i\in [N]:Z(U_j^i|Y_{1:T_c}^{1:N},U_{1:j-1}^{1:N},\\
&~~~~~~~~~~~~~~~~U_j^{1:i-1})\leq \delta_N\},
\end{aligned}
\end{equation}
with $U_{1:0}^{1:N}=\emptyset$. The multilevel encoding procedure goes as follows.
\begin{itemize}
	\item For the $j$th ($j\in[T_c]$) sub-channel, define the information bit set $\mathcal{I}_j$, frozen bit set $\mathcal{F}_j^r$ and deterministic bit set $\mathcal{F}_j^d$ according to (\ref{PC-I}), (\ref{PC-Fr}) and (\ref{PC-Fd}) respectively. 
	\item Insert information bits to $\{u_j^i\}_{i\in \mathcal{I}_j}$ and frozen bits to $\{u_j^i\}_{i\in \mathcal{F}_j^r}$, and compute $\{u_j^i\}_{i\in \mathcal{F}_j^d}$ according to (\ref{DetBit}).
	\item Compute $\mathbf{z}_j=\mathbf{u}_j\mathbf{G}_N$ for each $j\in[T_c]$ and generate the final coded frame by $\mathbf{X}=\big{(}\mathbf{z}_1;...;\mathbf{z}_{T_c}\big{)}$. 
\end{itemize}

The sender transmits $\mathbf{X}$ column by column. Having received $\mathbf{Y}$, the receiver performs multistage decoding.
In the $j$th ($1\leq j\leq T_c$) stage, the decoder decodes $\mathbf{u}_j$ with the aid of the estimates in previous stages:
	\begin{equation}
	\bar{u}_j^{i}=
	\begin{cases}
	u_j^{i},  \text{~~~~~~~~~~~~~~~~~~~~~~~~~~~~~~~~~~~~~~~~~if } i\in \mathcal{F}_j^r \\
	\arg\max_{u\in\{0,1\}}P_{U_j^{i}|U_j^{1:i-1}}(u|u_j^{1:i-1}),\text{~~if } i\in \mathcal{F}_j^d\\
	\arg\max_{u\in\{0,1\}}P_{U_j^{i}|Y^{1:N}_{1:T_c}U_{1:j-1}^{1:N}U_j^{1:i-1}}(u|y^{1:N}_{1:T_c},\\
	~~\bar{u}_{1:j-1}^{1:N},u_j^{1:i-1}),\text{~~~~~~~~~~~~~~~~~~~~~~~~~if } i\in \mathcal{I}_j
	\end{cases}
	\end{equation}
where $\bar{u}_{1:0}^{1:N}=\emptyset$.

The block error probability of the $j$th component polar code provided that the previous component codes are correctly decoded can be upper bounded by
\begin{equation}
P_e^{(j)} \leq\sum_{i\in\mathcal{I}_j}{Z(U_j^i|Y_{1:T_c}^{1:N},U_{1:j-1}^{1:N},U_j^{1:i-1})}=O(2^{-N^\beta})
\end{equation}
according to the definition of the information bit set. Thus, the overall frame error probability can be upper bounded by
\begin{equation}
\label{FER}
P_e \leq\sum_{j=1}^{T_c}P_e^{(j)}=T_cO(2^{-N^\beta}).
\end{equation}

The asymptotic rate of the $j$th ($j\in[T_c]$) component polar code is
\begin{equation}
\lim\limits_{N\rightarrow\infty}R_j=\lim\limits_{N\rightarrow\infty}\frac{|\mathcal{I}_j|}{N}=I(X_j;Y_{1:T_c}|X_{1:j-1}).
\end{equation} 
Thus, the asymptotic rate of the scheme is
\begin{align}
\lim\limits_{N\rightarrow\infty}R&=\lim\limits_{N\rightarrow\infty}\frac{1}{T_c}\sum_{j=1}^{T_c}R_j
=\frac{1}{T_c}I(X_{1:T_c};Y_{1:T_c}), \label{AR}
\end{align}
which equals the ergodic capacity of the block fading channel.

From (\ref{FER}) and (\ref{AR}) we can claim that our proposed polar coding scheme achieves the ergodic capacity of binary-input block fading channels with only CDI.

\subsection{Only CSI-R Available}
From the discussion in Section \ref{S:problemstate} we know that, to achieve capacity in the CSI-R case, it is sufficient to use a polar code for each sub-channel independently with the method of \cite{liu2016fading}. The benefit of independent coding is that different sub-channels can be decoded in parallel, which can improve the throughput of the communication system. Nevertheless, a joint encoding and decoding approach can significantly improve the finite length performance.

Consider a BICM-based approach that the transmitter uses an interleaver to randomize symbols in frame. The equivalent channel seen by the receiver after deinterleaving is an i.i.d. fading channel with known CSI-R if the interleaver is perfect. Thus, the scheme of \cite{liu2016fading} is applicable. If the coherent time is $T_c$ and the number of blocks in a frame is $N$, the total code length will be $T_cN$. Thus, to use standard polar codes, $T_cN$ should be the power of 2.

The frame error probability of the parallel scheme can be upper bounded by $$P_e^{Para}\leq T_cO(2^{-N^\beta}).$$
As a comparison, the frame error probability of the BICM-base scheme (assuming a perfect interleaver) satisfies $$P_e^{BICM}\leq O(2^{-(T_cN)^\beta}).$$ Obviously the finite length performance is significantly improved by interleaving, especially when $T_c$ is large. However, the decoding latency of the parallel scheme, assuming $T_c$ decoders working simultaneously, is only $1/(T_c(1+\log T_c/\log N))$ of that for the BICM-based scheme with the same frame size, since the decoding complexity of polar codes is $O(N\log N)$. This shows a trade-off between performance and throughput when used in practice. Since the scheme of \cite{liu2016fading} is capacity-achieving, we can claim that our proposed schemes in this subsection are also capacity-achieving.

\subsection{Full CSI Available}
When both the transmitter and the receiver have perfect CSI, we can simply use standard polar codes for each coherent interval if the coherent time is sufficiently large, as \cite{Santos2013rayleigh} has shown. Otherwise, we can use the CSI-R scheme introduced in the last subsection if the transmitter does not do power control, or design a CSI-R scheme for the equivalent channel after power allocation if it does.

\section{Numerical Results}
\label{S:Numerical}
In the CDI case, the average mutual information (per symbol) of a block fading channel is
\begin{equation}
\label{MI_CDI}
\frac{1}{T_c}I(X_{1:T_c};Y_{1:T_c})=\frac{1}{T_c}\sum_{\mathbf{x}}p(\mathbf{x})\int_{\mathbf{y}} p(\mathbf{y}|\mathbf{x})\log\frac{p(\mathbf{y}|\mathbf{x})}{p(\mathbf{y})}\mathrm{d}\mathbf{y},
\end{equation}
and the mutual information of the $j$th ($j\in[T_c]$) sub-channel is
\begin{equation}
\begin{aligned}
&~~~~I(X_j;Y_{1:T_c}|X_{1:j-1})\\
&=\sum_{x_{1:j}}p(x_{1:{j}})\int_{\mathbf{y}} p(\mathbf{y}|x_{1:{j}})\log\frac{p(\mathbf{y}|x_{1:{j}})}{p(\mathbf{y}|x_{1:{j-1}})}\mathrm{d}\mathbf{y}.
\end{aligned}
\end{equation}



As an example, we assume $h$ follows the Rayleigh distribution with PDF
\begin{equation}
f(h)=\frac{h}{\sigma_h^2}e^{-\frac{h^2}{2\sigma_h^2}}.
\end{equation}
Then the transition PDF of (\ref{transp}) is given by
\begin{equation}
p(\mathbf{y}|\mathbf{x})=\int_{0}^{\infty} \big{(}\prod_{j=1}^{T_c}\frac{1}{\sqrt{2\pi\sigma^2}}e^{-\frac{(y_j-hx_j)^2}{2\sigma^2}}\big{)}\frac{h}{\sigma_h^2}e^{-\frac{h^2}{2\sigma_h^2}}\mathrm{d}h.
\end{equation}

\begin{figure}[tb]
	\centering
	\includegraphics[width=8.5cm]{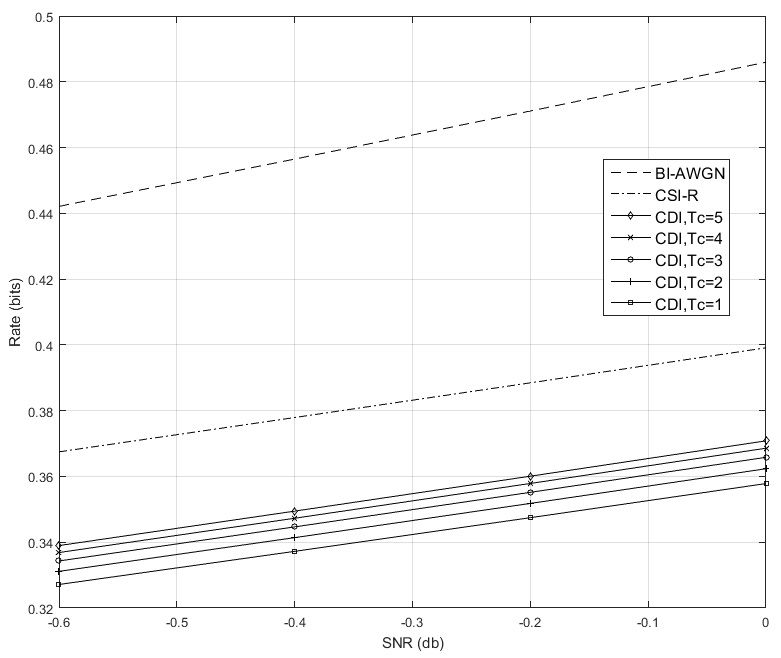}
	\caption{Achievable rates of binary-input block Rayleigh fading channels.} \label{fig:tc4}
\end{figure}
Since this channel is symmetric, we will only consider uniformly distributed channel inputs. Fig. \ref{fig:tc4} shows a comparison of achievable rates of binary-input AWGN channel, binary-input Rayleigh fading channel with CSI-R, and binary-input block Rayleigh fading channels of different coherent time with only CDI. We can see that as the coherent time increases, the achievable rate with only CDI gets closer and closer to that with CSI-R. It has been shown for several cases (e.g., \cite{peleg1998incoherent,liang2004noncoherent}) that the noncoherent capacity of a block fading channel will approach the coherent capacity as $T_c\rightarrow\infty$. In the considered SNR region (-0.6 db to 0 db), a binary-input block Rayleigh fading channel with 5 symbols per coherent interval has a performance gain about 0.25 db over the i.i.d. fading channel under the CDI assumption, and a performance loss about 0.5 db compared with the CSI-R curve.

\begin{figure}[tb]
	\centering
	\includegraphics[width=8.5cm]{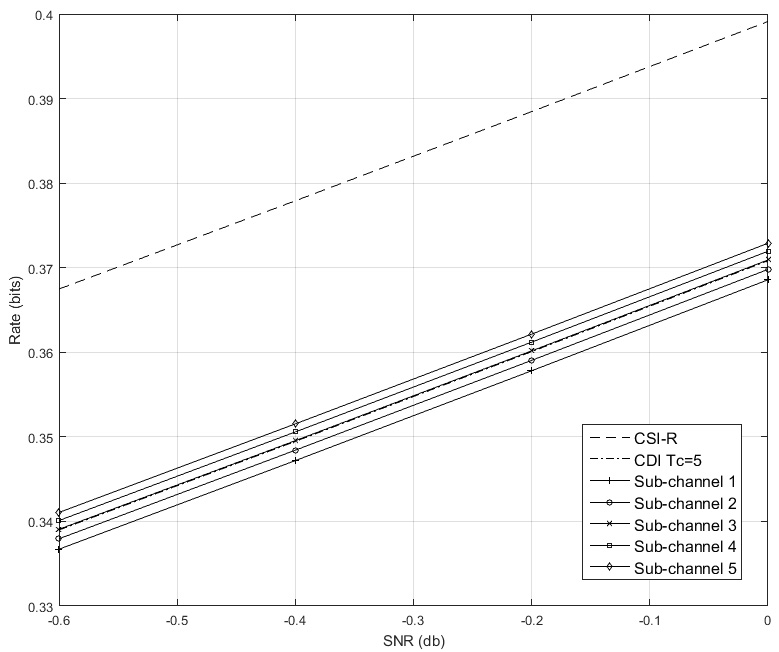}
	\caption{Capacity of sub-channels of a block Rayleigh fading channel with coherent time $T_c=5$.} \label{fig:sub4}
\end{figure}
Fig. \ref{fig:sub4} shows the achievable rates of five sub-channels of a block Rayleigh fading channel with coherent time $T_c=5$, compared with that of the original channel and the CSI-R rate. We can see that the achievable rate of a sub-channel increases with its index. This can be intuitively  explained as follows. After decoding a sub-channel, the decoder gains more knowledge about the CSI (although not explicitly shown), and the achievable rates of the following sub-channels become larger. Thus, our proposed CDI scheme can be seen as a realization of the joint channel estimation and data transmission paradigm in noncoherent communications.

\section{Discussion}
\label{S:discuss}
In this paper, we take a coded modulation approach to solve the problem of coding for block fading channels. By viewing transmitted symbols in a coherent block as a supersymbol, we design a multilevel polar coding scheme when only CDI is available, and a parallel scheme as well as a BICM-based scheme when instantaneous CSI is available at the receiver or at both sides. All of these schemes are capacity-achieving. 

It is known that in coded modulation, BICM can not achieve coded modulation capacity, because it treats all bit levels equally, which in fact may have some dependency. Similarly, in block fading channels, if we design a BICM-based scheme for the CDI case, it will fail to achieve the capacity since it erases the relation between channel uses in the same coherent block. However, if the receiver knows the instantaneous CSI, each channel use will be indeed independent from others from the receiver's perspective, and the BICM-based scheme can also achieve capacity.


Another approach to deal with the problem of coding for block fading channels is the multiple access channel (MAC) approach, which views a block fading channel of coherent time $T_c$ as a $T_c$-user MAC. We will briefly explain the connection between this approach and ours. From the MAC perspective, code design will be based on MAC polar codes (e.g., \cite{arikan2012sw}). Our proposed MLC-based scheme can be seen as a special case of the MAC-based scheme, i.e., it is equivalent to a MAC polar code designed to achieve a corner point of the achievable rate region of the $T_c$-user MAC. By using other permutations for MAC polarization, one can allocate rates for different "users" more flexibly. However, although this connection is valid, the MAC approach looks unnecessarily complicated for our problem. In the end, it's chain rule, not MAC, that matters for our problem.



Although we have only considered binary inputs in this paper, the proposed schemes can be readily applied to non-binary cases. The idea of this work may also be extended to fading channels with memory. We will leave these to our future work.



\bibliographystyle{IEEEtran}
\bibliography{Polar_BlockFading}

\end{document}